# Multiferroic and magnetoelectric nature of GaFeO$_3$, AlFeO$_3$ and related oxides

**Rana Saha[a], Ajmala Shireen[b], Sharmila N. Shirodkar[c], Umesh V. Waghmare[c,d], A. Sundaresan[a,b,d] and C. N. R. Rao[a,b,d,*]**

[a]*Chemistry and Physics of Materials Unit, Jawaharlal Nehru Centre for Advanced Scientific Research, Jakkur P.O., Bangalore 560064, India.*

[b]*New Chemistry Unit, Jawaharlal Nehru Centre for Advanced Scientific Research, Jakkur P.O., Bangalore 560064, India.*

[c]*Theoretical Science Unit, Jawaharlal Nehru Centre for Advanced Scientific Research, Jakkur P.O., Bangalore 560064, India.*

[d]*International Centre for Materials Science, Jawaharlal Nehru Centre for Advanced Scientific Research, Jakkur P.O., Bangalore 560064, India.*

*Fax: +91-80-22082766; Tel: +91-80-22082761.*

*E-mail: cnrrao@jncasr.ac.in.*

**Abstract**

GaFeO$_3$, AlFeO$_3$ and related oxides are ferrimagnetic exhibiting magnetodielectric effect. There has been no evidence to date for ferroelectricity and hence multiferroicity in these oxides. We have investigated these oxides as well as oxides of the composition Al$_{1-x-y}$Ga$_x$Fe$_{1+y}$O$_3$ (x = 0.2, y = 0.2) for possible ferroelectricity by carrying out pyroelectric measurements. These measurements establish the occurrence of ferroelectricity at low temperatures below the Nèel temperature in these oxides. They also exhibit significant magnetoelectric effect. We have tried to understand the origin of ferroelectricity based on non-centrosymmetric magnetic ordering and disorder by carrying out first-principles calculations.

1. Introduction

GaFeO$_3$ crystallizes in the non-centrosymmetric orthorhombic structure (*Pna*2$_1$) while α-Fe$_2$O$_3$ crystallizes in the corundum structure ($R\bar{3}c$), and Ga$_2$O$_3$ has the stable monoclinic structure [1]. AlFeO$_3$ also crystallizes in the orthorhombic crystal structure with the *Pna*2$_1$ space group [2-4], although both the parent binary oxides, Al$_2$O$_3$ and Fe$_2$O$_3$, crystallize in the corundum structure ($R\bar{3}c$). Crystal structures of AlFeO$_3$ and GaFeO$_3$ are made up of alternative layers of cations (Fe/Ga/Al) and oxygen ions stacked along the crystallographic *c* direction. There are four cationic sites present in GaFeO$_3$ and AlFeO$_3$, with Fe1, Fe2 and Al2 (Ga2) in the octahedral environment. The octahedra share edges whereas the tetrahedra Al1 (Ga1) share oxygen at the corners. There is site disorder among the cationic sites and site occupancies do not vary with lowering of temperature [5]. GaFeO$_3$ and AlFeO$_3$ possess collinear ferrimagnetic structure [6] with Néel temperature ($T_N$) of 210 and 250 K respectively. Magnetic ordering occurs due to cation-oxygen-cation superexchange antiferromagnetic interactions. Disorder plays a significant role in determining the magnetic and dielectric properties of GaFeO$_3$ and AlFeO$_3$. There are indications in the earlier literature that both GaFeO$_3$ and AlFeO$_3$ are piezoelectric [7, 8], and a recent study reports magnetodielectric effect in these oxides [9]. What is not clear, however, is whether GaFeO$_3$ and AlFeO$_3$ are ferroelectric, and if so whether they are multiferroic. We have investigated GaFeO$_3$, AlFeO$_3$ and a few compositions of general formula Al$_{1-x-y}$Ga$_x$Fe$_{1+y}$O$_3$ (x + y < 1) for possible ferroelectricity by carrying out careful pyroelectric current measurements, since dielectric hysteresis measurements are unreliable in such leaky materials [10]. The pyroelectric measurements have indeed established the occurrence of ferroelectricity below $T_N$ and thereby the multiferroic nature of these oxides. We have carried out first-principles calculations to understand the origin of ferroelectricity in these simple iron oxides.

## 2. Experimental Details

$Al_{1-x}Ga_xFeO_3$ (x = 0.0, 0.5, 1.0) and $Al_{1-x-y}Ga_xFe_{1+y}O_3$ (x = 0.9, y = 0.1; x = 0.8, y = 0.2 and x = 0.2, y = 0.2) were prepared by conventional solid state synthesis. Stoichiometric amounts of the precursor oxides namely $Fe_2O_3$, $Al_2O_3$ and $Ga_2O_3$ were ground repeatedly to a fine mixture and was heated to a sintering temperature of $1400^0$ C through intermediate heating. Heating duration was varied from sample to sample till a phase pure compound was obtained in each case. The X-ray diffraction patterns for all the oxides are shown in Fig. 1. Lattice parameters of these oxides are given in Table 1.

Pyroelectric current measurements on $GaFeO_3$ and $AlFeO_3$ were carried out using a Keithley 6517A electrometer. The samples were poled down to lowest temperature by applying an electric field and shorted after removing the field. Pyroelectric current was measured while heating the samples at the rate of 4 K/min. To measure the leakage current, the sample was cooled down to the lowest temperature and then an electric field of the same magnitude as poling electric field was applied for the same duration as the poling time. Before the leakage measurement, we removed the electric field and waited for some time without any shorting of the electrodes. Leakage current data were recorded during heating the sample at the rate of 4 K/min. The measured leakage current was subtracted from the pyroelectric current. This current reverses direction on positive and negative poling. Polarization data were obtained by integrating the current and plotted against temperature.

## 3. Results and Discussion

The pyroelectric current in $AlFeO_3$ and $GaFeO_3$ show peaks ($T_p$) around 100 and 90 K respectively (insets of Fig. 2). Pyroelectric behavior of $AlFeO_3$ and $GaFeO_3$ was similar but polarization value of $AlFeO_3$ was found to be lower than that of $GaFeO_3$ (Fig.2). $AlFeO_3$, $GaFeO_3$ and related oxides can be quite leaky, but the measured leaky currents were generally small and constant with a small anomaly near $T_p$. In Fig. 3, we compare the magnitude of the leaky current in the case of $AlFeO_3$ to illustrate this aspect. The leakage contribution was deducted from the pyroelectric current. To further confirm the ferroelectric nature of the samples, we performed the polarization reversal experiment in GaFeO$_3$ following Kimura *et al.* [11]. The sample was poled down to 10 K with an electric field of + 2 kV/cm. After shorting the electrodes to remove the surface charge, an opposite electric field of – 6.25 kV/cm was applied at 10 K. The electric field was then turned off and the electrodes were shorted again. Pyrocurrent was measured while heating the sample at a rate of 4 K/min in absence of any electric field but we did not observe the switching of polarization to opposite polarity indicating that either the coercieve field is still very high or the applied bias is not sufficient to reverse the direction of dipoles. To overcome this, we applied the reverse bias (-6.25 kV/cm) to the poling bias at a higher temperature (60 K) but well below the ferroelectric transition temperature after poling the sample in presence of +2 kV/cm. We could observe the switching of the polarization to opposite polarity indicating the dipolar nature of GaFeO$_3$. We find a significant effect of magnetic fields on the polarization of AlFeO$_3$ and GaFeO$_3$ as shown in Figs. 2a and 2b respectively. Suppression of polarization by magnetic fields indicates significant magnetoelectric coupling. Neutron diffraction measurements [5] Al$_{1-x}$Ga$_x$FeO$_3$ (x = 0.0, 0.5, 1.0) have shown that the octahedral sites are highly distorted due to the mixed occupancy of cations with different cationic radii. Furthermore, the unit cell parameters exhibit broad minima in the 100 K range where these oxides show pyroelectric current peaks.

We have studied the magnetic properties of oxides of the general composition Al$_{1-x-y}$Ga$_x$Fe$_{1+y}$O$_3$ (x + y < 1) with excess Fe in order to increase the $T_N$ towards room temperature. All the oxides with excess Fe also retain the orthorhombic *Pna*2$_1$ structure (Table 1). In Fig. 4, we show typical magnetization data of such composition. In Al$_{1-x-y}$Ga$_x$Fe$_{1+y}$O$_3$ with x = 0.9, y = 0.1 and x = 0.8, y = 0.2 the $T_N$ values are 245 and 290 K respectively (Table 1). In order to improve $T_N$ further, we have examined properties of compositions Al$_{1-x-y}$Ga$_x$Fe$_{1+y}$O$_3$ (x = 0.2, y = 0.2) ($T_N$ = 325 K). All these oxides show ferroelectric behavior, with pyroelectric current maxima in the 90 K region. Al$_{0.5}$Ga$_{0.5}$FeO$_3$, shows a pyrocurrent maximum around 95 K as shown in Fig.5 with the polarization value comparable to that of GaFeO$_3$ and with significant magnetoelectric effect.

The results presented above establish the ferroelectric nature and hence the multiferroic nature of GaFeO$_3$, AlFeO$_3$ and related oxides. The multiferroic nature in these oxides is clearly related to the magnetic properties of these materials. There are a few recent reports where ferroelectricity in certain iron oxides directly arises from magnetic interactions. Thus, in GdFeO$_3$ ferroelectric polarization is induced by striction through exchange interaction between the Gd and Fe spins [12]. Ferroelectricity in YFe$_{1-x}$Mn$_x$O$_3$ arises from magnetic interactions involving the two transition metal ions [13]. The oxides in the present study have only one magnetic ion (Fe$^{3+}$), but there is considerable disorder which makes the octahedral sites magnetically non-equivalent.

We have carried out first-principles calculations based on density functional theory (DFT) with a spin-density dependent exchange correlation energy functional of a generalized gradient approximated (GGA) (PerdewWang 91 (PW 91)) form [14] as implemented in the Vienna *ab* initio Simulation Package (VASP) [15], [16]. The projector augmented wave (PAW) method [17] was used to capture interaction between ionic cores and valence electrons. An energy cut off of 400 eV was used for the plane wave basis and integrations over the Brillouin were carried out using a regular 3x2x2 mesh of k-points. The structure was optimized to minimum energy using Hellman-Feynman forces, while maintaining the lattice constants at their experimental values. Minimum energy states with different magnetic ordering were obtained through appropriate initialization of the spins on Fe sites, simulating a unit cell containing 8 formula units (f.u.) of GaFeO$_3$ i.e. a unit cell of 40 atoms.

We present an argument based on symmetry and disorder to explain a possible origin of the observed ferroelectricity. GaFeO$_3$ belongs to an orthorhombic structure with a non-centrosymmetric space group (*Pna*2$_1$). In this structure, polarization along *z*-axis can be non-zero because the inversion symmetry is broken and there is no symmetry that involves a reflection in *xy* plane. We note that *P$_z$* is not switchable (cannot be reversed on the application of electric field of opposite polarity). Anti-site disorder between Fe and octahedral Ga sites is known to be present in the experimental samples. In the absence of such disorder, magnetic ordering of GaFeO$_3$ in its ground state is antiferromagnetic (AFM); Fe at Fe1 and Fe2 sites have antiparallel spins. We note that the effective interaction among Fe at Fe1 sites (or Fe at Fe2 sites) is ferromagnetic, whereas that between Fe at Fe1 and Fe2 sites is antiferromagnetic. This is expected to lead to magnetic frustration in GaFeO$_3$. One of the symmetry operations of *Pna*2$_1$, ($\bar{x}$, $\bar{y}$, $z$+1/2), transforms a pair of Fe1 sites to the other of Fe1 sites. As the magnetic moments of Fe at all the Fe1 sites are equal and opposite of those at Fe2 sites (Table 2 and Fig. 6a), the magnetic structure preserves $C_{2z}$ rotational symmetry. Thus, it has the planar inversion symmetry ($x, y, z$) in the $xy$ plane, and cannot have $P_x$, $P_y \neq 0$.

The anti-site disorder exists between Fe and Ga at the Fe2 and Ga2 sites due to the nearly same ionic radii of Fe$^{3+}$ (0.645 Å) and Ga$^{3+}$ (0.62 Å) in octahedral coordination [5]. We have simulated this disorder by interchanging one of Fe at Fe2 site with one Ga at Ga2 site in a single unit cell (which corresponds to 12.5 % of anti-site defects) in the AFM state (where Fe at Fe1 and Ga2/Fe2 have antiparallel spins). Though the ordered state is energetically more favourable than the disordered state, the difference in their energies is only 36 meV/f.u.[5] which explains the high occurrence of this anti-site defect. Due to this anti-site defect with Ga at Fe2 site and magnetic frustration, magnetic moments on Fe ions at the four Fe1 sites no longer remain equal and the same holds for Fe at Fe2 sites (Table 2 and Fig. 6b). Such magnetic order breaks the inversion symmetry in the '*ab*' plane and permits a non-zero polarization in the '*x*' and '*y*' directions.

We note that the inversion symmetry is broken due to anti-site disorder even in the absence of magnetic ordering in the system. This leads to non-zero polarization in the '*x*' and '*y*' directions, but the polarization induced only by the anti-site disorder is not switchable. Al$_{1-x}$Ga$_x$FeO$_3$ (x = 0.0, 0.5, 1.0) compounds exhibit a strong spin-phonon coupling [5], which is manifested in anomalies in their Raman spectra close to magnetic transitions [18]. Due to the unequal magnetic moments developed on the Fe ions in the presence of anti-site defects, unequal forces are exerted on the ions which lead to small structural distortions. These non-centrosymmetric structural distortions due to spin-phonon coupling are switchable with field and, induce switchable non-zero $P_x$ and $P_y$. The phonon ($\mu$) that couples very strongly with the spin, involves oxygen displacements and softens with decreasing temperature [18]. It has a maximum overlap with the first order spin-phonon coupling ($\Delta$), as reflected in the forces on atoms due to change in magnetic ordering (refer to Figure 6(c)).

We now give a phenomenological theory to explain the observed polarization arising from magnetic order and spin-phonon coupling. Let $\omega_\mu$, $Z^*_\mu$ and $u_\mu$ denote the frequency, mode effective charge and displacements of the ions associated with $\mu$ which couples to the spin. $M$ and $E$ are the magnetic order parameter and applied electric field respectively.

The polarization associated with structural distortion is given by:

$$P = Z^*_\mu u_\mu = \frac{Z^{*2}_\mu E}{\omega_\mu(T)^2} + \frac{\Delta Z^*_\mu}{\omega_\mu(T)^2} M(T)$$

The first and second terms in the above equation are the dielectric and spin-phonon coupling contributions to polarization respectively. Within Landau theory:

$$M(T) \propto (T_N - T)^{1/2}$$

and

$$\omega_\mu(T)^2 \propto |(T_P - T)|$$

Here, $T_N$ and $T_p$ are the Nèel temperature and the peak temperature in pyroelectric current respectively. This dependence of polarization on magnetization, spin-phonon coupling and phonon frequency gives a non-zero value of polarization at temperatures below $T_N$ and above the temperature of peak pyroelectric current (refer to Figure 6(d)). Such magnetostriction is expected to give rise to observable electromagnons [19]. To estimate this, we carried out first-principles calculations of the structure with the magnetic configuration obtained with time-reversed spins. We find that changes in structure involve rather small displacements of ions (~ $10^{-3}$ Å) and are within the errors in DFT calculations indicating a weak ionic contribution to magneto-electric coupling.

From the temperature dependent polarization estimated from pyroelectric current measurements, we note that it is nonzero well above $T_p$, and the magnitude of the pyroelectric current even 60 K above $T_p$ is comparable to its values at low temperatures. Thus, a pyroelectric current that is switchable with electric field, shows a broken inversion symmetry well above $T_p$. Within the model which we use to explain the observed results, a non-zero polarization arises from the broken inversion symmetry associated with magnetic ordering that occurs at a higher temperature ($T_N$). However, the magnitude of such polarization is rather small [20], of the order of 0.05 μC/cm$^2$, and is even smaller in the present case due to partial cancellation due to disorder. Once the magnetic order becomes sufficiently strong, phonons make an additional contribution to polarization (and pyroelectric current), giving rise to a feature of a maximum in the pyrocurrent at $T_p$. In this sense, (Al,Ga)FeO$_3$ seems to go through another improper transition at lower temperatures. While leakage current is likely to affect the pyroelectric current measurements on polycrystalline samples, it would have increased with temperature. This is not observed in the measurements thus supporting our model of the origin of polar order.

The pyroelectric measurements of spontaneous polarization can be used to obtain an insight into the main contributors to ferroelectricity. These measurements of $P_s$ in GaFeO$_3$, shows an asymmetry with respect to the poling field applied during sample cooling. Unlike $P_x$ and $P_y$, the sign of the pyroelectric current due to non-switchable $P_z$ does not change on reversing the poling field. Hence, the spontaneous polarization which has both switchable ($P_x$ and $P_y$) and non-switchable ($P_z$) contributions, exhibits this asymmetry. The difference in the total polarization of positively poled and negatively poled samples give an estimate of contribution from pyroelectric change in $P_z$ which is ~ 0.075 μC/cm$^2$. Since the remainder of the polarization comes from $P_x$ and $P_y$, our observation of switchable spontaneous polarization in GaFeO$_3$ is quite robust. Thus, ferroelectricity in GaFeO$_3$ is a result of non-centrosymmetric magnetic ordering which arises from inherent magnetic frustration, and changes in it due to anti-site disorder as well as spin-phonon coupling.

## 4. Conclusions

In conclusion, the present study indicate that GaFeO$_3$, AlFeO$_3$ and related oxides of the general formula Al$_{1-x-y}$Ga$_x$Fe$_{1+y}$O$_3$ (x + y < 1) are ferroelectric below $T_N$ and also exhibit magnetoelectric effect. It is noteworthy that the values of polarization in these oxides are quite significant. Observation of multiferroic and magnetoelectric properties in simple oxides such as Al(Ga)FeO$_3$ is significant and suggests new venues for designing and understanding these fascinating properties.

**Figure Captions**

Fig.1. X-ray diffraction patterns of (a) GaFeO$_3$ (b) Al$_{0.5}$Ga$_{0.5}$FeO$_3$ (c) AlFeO$_3$ (d) Ga$_{0.9}$Fe$_{1.1}$O$_3$ (e) Ga$_{0.8}$Fe$_{1.2}$O$_3$ and (f) Al$_{0.6}$Ga$_{0.2}$Fe$_{1.2}$O$_3$.

Fig.2. Variation of electric polarization (*P*) as a function of temperature at + ve and – ve poling for (a) AlFeO$_3$ and (b) GaFeO$_3$ (after leakage subtraction) along with the effect of a 4 T magnetic field.

Fig.3. Temperature variation of (a) pyroelectric current (after subtraction); (b) leakage current and (c) pyroelectric current (before subtraction) for AlFeO$_3$ at -1 kV/cm.

Fig.4. Variation of magnetization (*M*) data under field-cooled (FC) zero-field-cooled (ZFC) conditions at 100 Oe as a function of temperature for (a) Ga$_{0.8}$Fe$_{1.2}$O$_3$ and (b) Al$_{0.6}$Ga$_{0.2}$Fe$_{1.2}$O$_3$. Inset shows the *M* vs *H* plot at 10 K.

Fig.5. Variation of electric polarization (*P*) as a function of temperature at + ve and –ve poling for Al$_{0.5}$Ga$_{0.5}$FeO$_3$ (after leakage subtraction) along with the effect of a 4 T magnetic field.

Fig.6. Magnetic structure of GaFeO$_3$, (a) without anti-site defect and (b) with anti-site defect. Red represents Fe1 sites and blue Fe2 sites. (a) In the ordered state, magnetic moments on Fe1 sites are equal and opposite to Fe2 sites and hence they do not break inversion symmetry in '*ab*' plane. (b) Due to anti-site defect, one of the Fe at Fe2 site is displaced to the Ga2 site. Also the magnetic moments at the four Fe1 or Fe2 sites no longer remain equal. The inversion symmetry in '*ab*' plane is broken which permits spontaneous polarization in the same plane. (c) Forces acting on the ions when the system goes from AFM to paramagnetic state. Forces are proportional to the first order spin-phonon coupling (Δ) responsible for non-zero spontaneous polarization in Fe(Al,Ga)O$_3$. (d) Variation of polarization with temperature within our spin-phonon coupling theory. The temperature variation of magnetization and frequency were approximated from the Landau theory of phase transitions. It is evident that the system shows non-zero polarization below $T_N$ in the presence of disorder.

**Table 1**

**Lattice parameters and magnetic properties of GaFeO$_3$, AlFeO$_3$ and related oxides**

| Compound formula | $a$ (Å) | $b$ (Å) | $c$ (Å) | $T_N$ (K) |
|---|---|---|---|---|
| GaFeO$_3$ | 5.0814 (2) | 8.7436 (3) | 9.3910 (2) | 210 |
| Al$_{0.5}$Ga$_{0.5}$FeO$_3$ | 5.0306 (1) | 8.6461 (2) | 9.3175 (2) | 220 |
| AlFeO$_3$ | 4.9806 (3) | 8.5511 (6) | 9.2403 (6) | 250 |
| Ga$_{0.9}$Fe$_{1.1}$O$_3$ | 5.0811 (1) | 8.7444 (1) | 9.3924 (1) | 245 |
| Ga$_{0.8}$Fe$_{1.2}$O$_3$ | 5.0834 (1) | 8.7521 (1) | 9.4032 (1) | 290 |
| Al$_{0.6}$Ga$_{0.2}$Fe$_{1.2}$O$_3$ | 5.0299 (1) | 8.6425 (1) | 9.3219 (1) | 325 |

**Table 2: Magnetic moments on individual Fe ions ($\mu_B$) in the ordered and disordered AFM state**

| Atom no. | Cation site | Magnetic moment ($\mu_B$) | |
|:---:|:---:|:---:|:---:|
| | | Ordered | Disordered |
| 1 | Fe1 | 3.62 | 3.71 |
| 2 | Fe1 | 3.62 | 3.51 |
| 3 | Fe1 | 3.62 | 3.70 |
| 4 | Fe1 | 3.62 | 3.62 |
| 5 | Fe2/Ga2 | -3.59 | -3.75 |
| 6 | Fe2 | -3.59 | -3.62 |
| 7 | Fe2 | -3.59 | -3.61 |
| 8 | Fe2 | -3.59 | -3.59 |

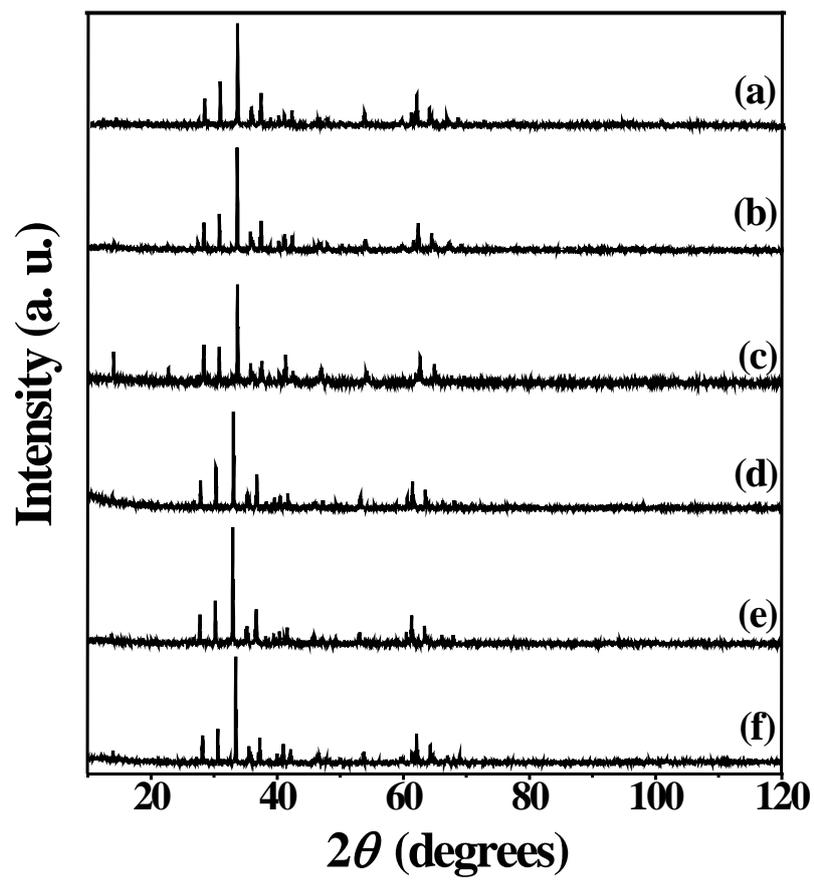

**Fig. 1**

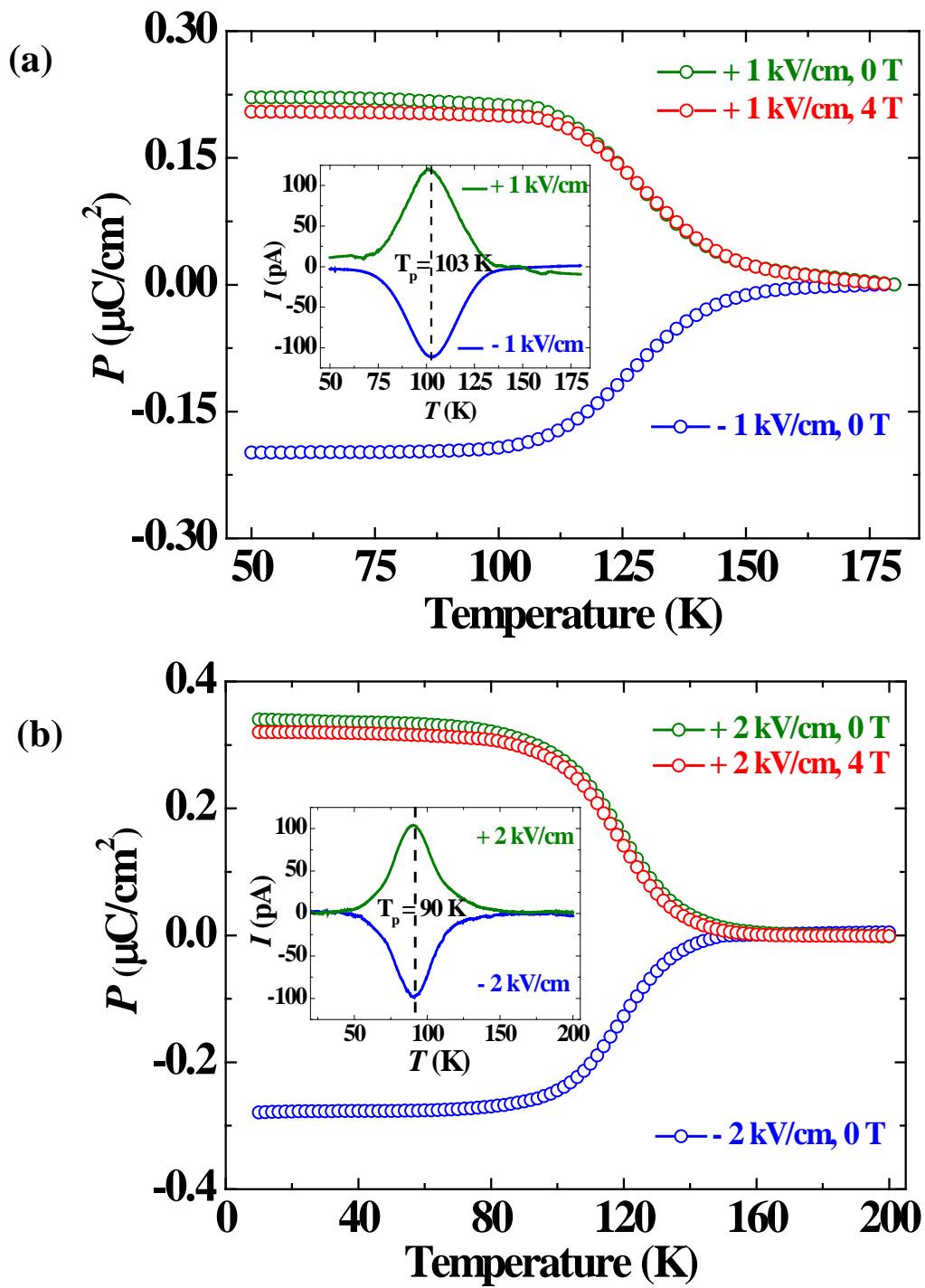

**Fig. 2**

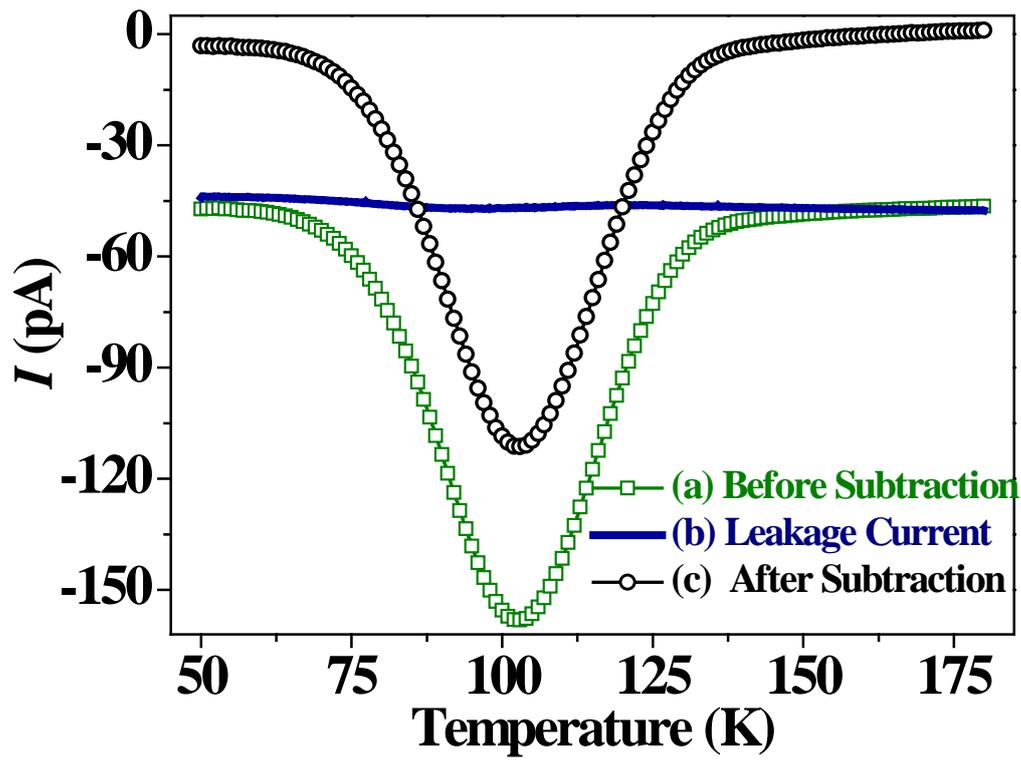

Fig. 3

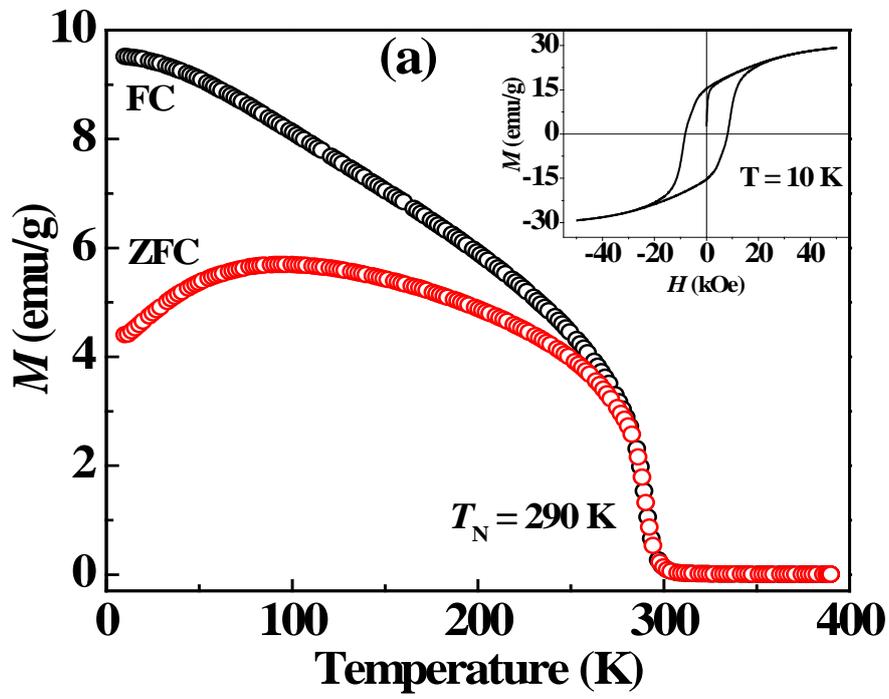
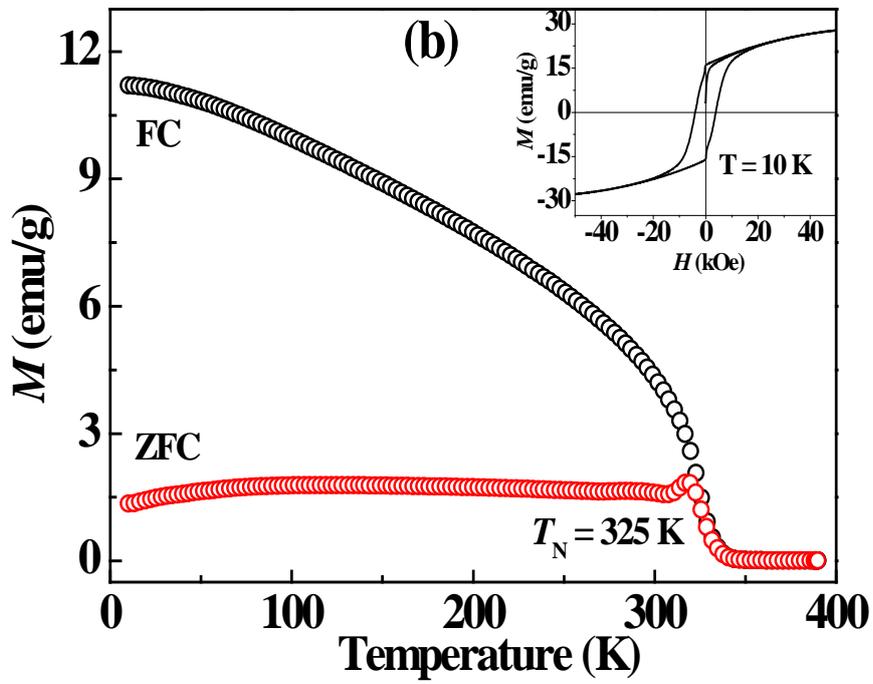

**Fig. 4**

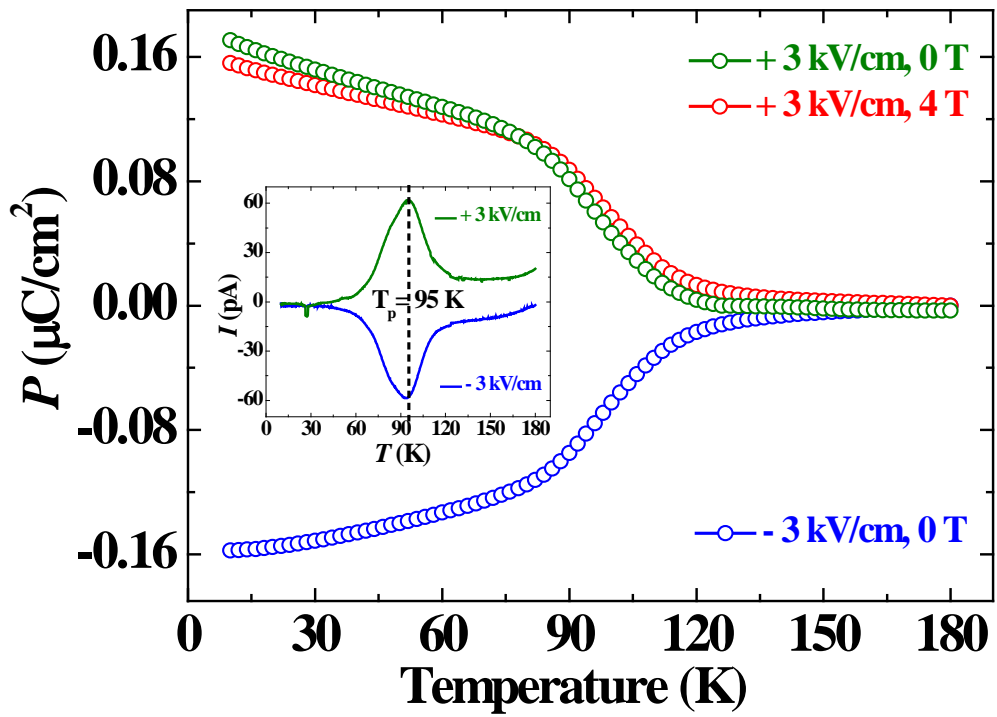

**Fig. 5**

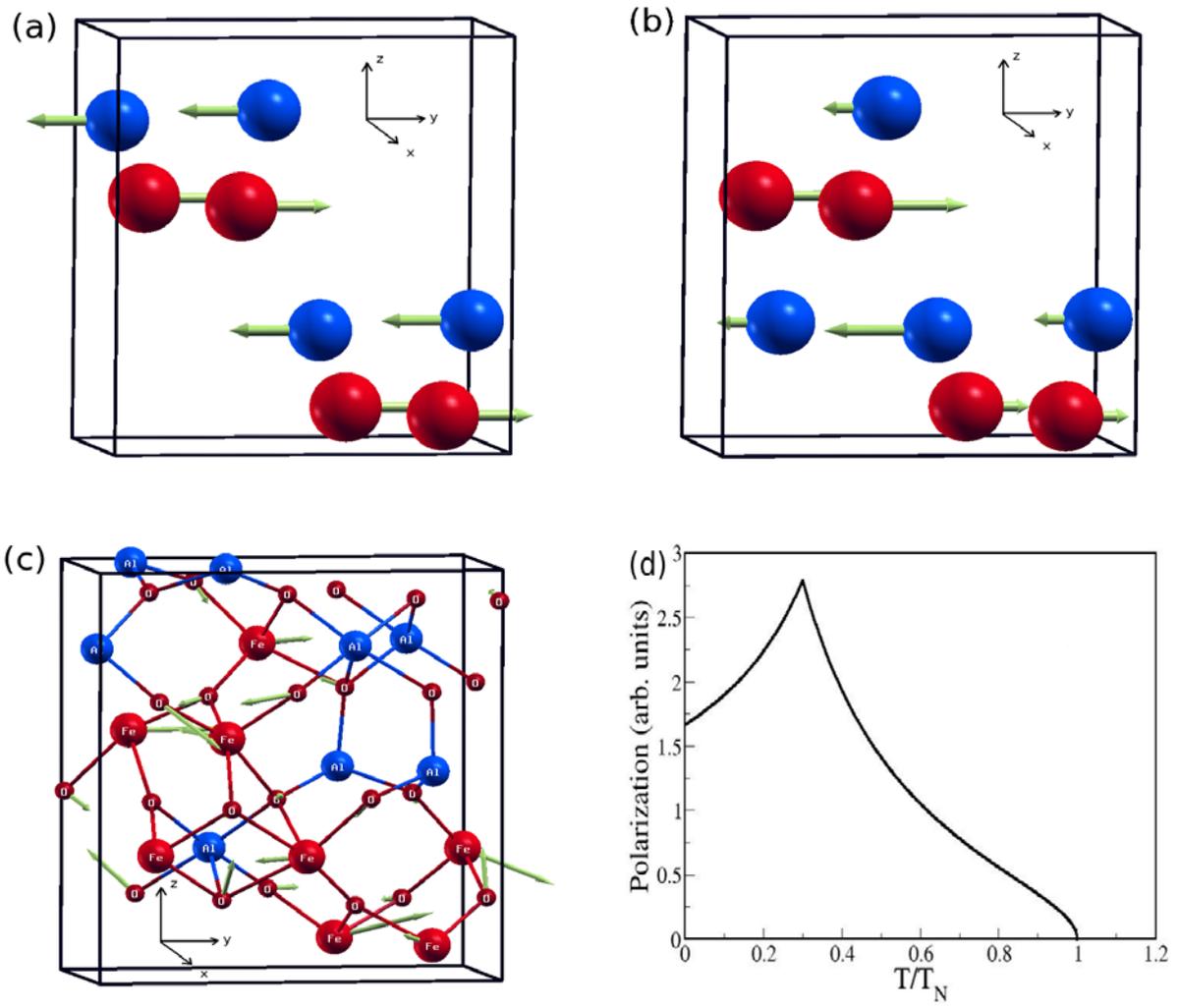

**Fig. 6**

## Graphical Abstract

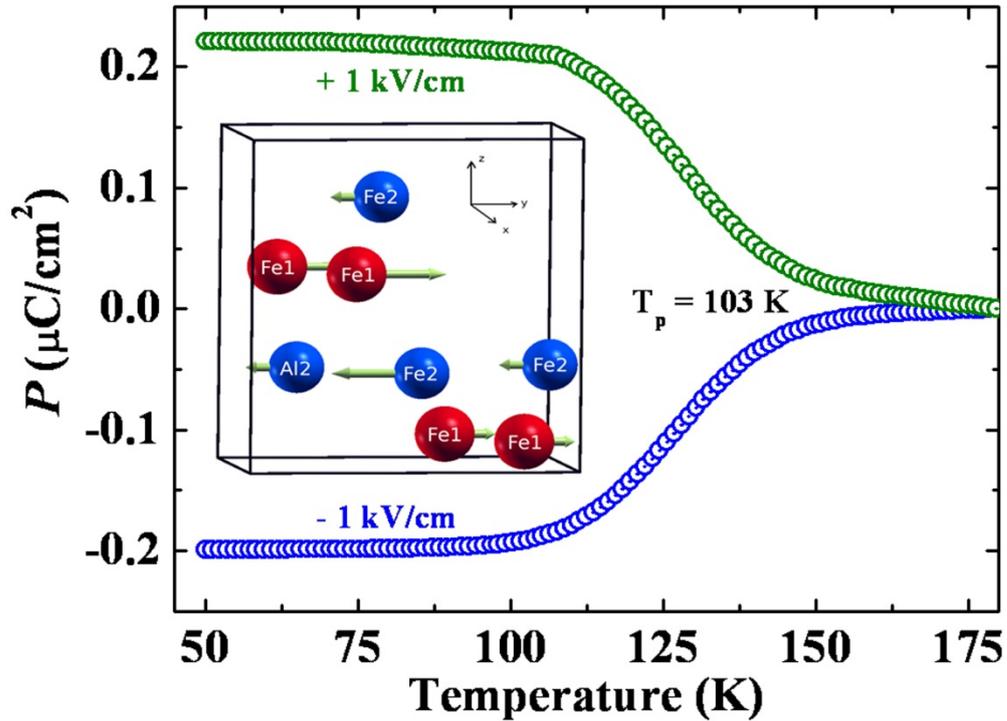

Occurence of ferroelectricity in GaFeO$_3$, AlFeO$_3$ and related oxides is indicated by pyroelectric measurements. These oxides are ferrimagnetic and also exhibit magnetoelectric effect. Theoretical studies provide an insight into the origin of ferroelectricity in these oxides.